\documentclass[cits]{PoS}
\usepackage{array}
\usepackage{dsfont}
\newcolumntype{S}{>{\centering\arraybackslash} m{.4\linewidth} }
\title{Investigation of the {$\eta'$-$\eta_c$}-mixing with improved stochastic estimators}

\ShortTitle{{$\eta'$-$\eta_c$}-mixing with improved stochastic estimators}

\author{\speaker{Christian~Ehmann} and Gunnar S.\ Bali\\
        Institut f\"ur Theoretische Physik, Universit\"at Regensburg,\\
 93040 Regensburg, Germany\\
        E-mail: \email{christian.ehmann@physik.uni-regensburg.de},
\email{gunnar.bali@physik.uni-regensburg.de}}

\abstract{Charmonia are flavour singlet mesons and thus in principle
contributions from disconnected quark line diagrams might affect their
masses, either directly or via mixing with other flavour singlet
channels. We present a first study that takes both effects into
account. 
We employ improved stochastic all-to-all propagator techniques
(including new methods) to calculate the diagrams that appear within
the mixing matrix between the $\eta'$ and the $\eta_c$.
The runs are initially performed on $N_f=2$
$16^3\times 32$ configurations with the non-perturbatively
improved Sheikholeslami-Wilson action, both for valence and sea quarks.}

\FullConference{The XXVI International Symposium on Lattice Field Theory\\
		 July 14-19 2008\\
		 Williamsburg, Virginia, USA}

\begin{document}

\section{Introduction}
Lattice calculations of the
charmonium $S$-wave hyperfine splitting
tend to underestimate the experimental value of
117~MeV~\cite{McNeile:2004wu,deForcrand:2004ia,Levkova:2008qr,Gottlieb:2005me,Ehmann:2007hj,Namekawa:2008ft}.
However, in none of these studies all systematics have been
addressed as yet. One such source of error is the difference
of the running of the QCD coupling between low momenta $\simeq mv^2$
that determine the
spin averaged splittings and the high momenta $> mv$ responsible
for the fine structure.
Too few or too heavy sea quark flavours will result in a comparably smaller
coupling at high momenta, resulting in such an
underestimation~\cite{Bali:1998pi}. Furthermore, the finestructure is
particularly sensitive to short distance physics, necessitating
a careful extrapolation to the continuum limit.

Here we will investigate the effect of quark-antiquark
annihilation diagrams,
the neglection of which
might be an additional source of the discrepancy. Although
heavy, compared to the chiral symmetry breaking scale, the $\eta_c$ state
might still sense some remnant axial $U(1)$  effect that could result in
an upward mass shift, perhaps of a few MeV. Another aspect
is the mixing with other flavour singlet states, e.g.\ the
pseudoscalar glueball or the $\eta'$ meson and its radial excitations.

We study both effects: quark annihilation diagrams~\cite{McNeile:2004wu,deForcrand:2004ia,Levkova:2008qr} and mixing
with the $\eta'$, hence called $\eta$ since we have $N_f=2$.
To investigate this mixing we construct a correlation matrix
containing both light and charm quark operators, the calculation
of which requires all-to-all-propagators.
The eigenvalues and appropriately normalized eigenvectors of this
matrix are indicative
of the magnitude of mixing between the two sectors.

\section{Simulation details}
Our runs are performed on $N_f=2$ $16^3\times 32$ configurations
generated by the QCDSF collaboration \cite{AliKhan:2003cu},
with a lattice spacing $a\approx 0.1145\,$fm, obtained from the
chirally extrapolated nucleon mass. For valence
and sea quarks we use the clover action, with $c_{sw}$
determined non-perturbatively. The charm quark mass was set
by tuning the spin-averaged charmonium
mass $\frac{1}{4}(m_{\eta_c}+3m_{J/\Psi})$
to its experimental value.
The pion on these lattices is rather heavy: $m_{\pi}=1007(2)\,$MeV.
However, this reduces the mass gap between the $\eta$ and the $\eta_c$
which might enhance mixing effects.
Computations took place on the local QCDOC using the Chroma software
library \cite{Edwards:2004sx}. 
We utilized
100 effectively de-correlated configurations, where on each
200 independent stochastic estimates were created.

\section{All-to-all propagator techniques}
In order to calculate the quark-antiquark annihilation diagrams, we
require propagators from any start to any end point.
An exact inversion of the Dirac Operator $M$ is not feasible
in terms of the memory requirements and computation time.
Hence we calculate
unbiased stochastic estimates of these all-to-all propagators.
We define random noise vectors $|\eta^i\rangle$,
$i=1,\ldots, N$ with components,
 \begin{equation}
 \eta^i_{\alpha,a,x} \quad = \quad \frac{1}{\sqrt{2}} (v+iw)\,,  \qquad v,w \in\{\pm1\}\,.
 \end{equation}
We solve the linear problem for these $N$ sources:
\begin{equation}
|s^i\rangle = M^{-1} |\eta^i\rangle\,,
\end{equation}
where,
\begin{equation}
 \frac{1}{N} \sum_i \eta^i_{\alpha,a,x} \eta^{i\,*}_{\beta,b,y} = \delta_{x,y}\delta_{a,b}\delta_{\alpha,\beta} + O\left(\frac{1}{\sqrt{N}}\right)\,.
\end{equation}
The propagator $M^{-1}$ can now be estimated as follows:
\begin{equation}
\overline{|s\rangle\langle\eta|}\,\,:=\,\,\frac1N
\sum_i |s^i\rangle\langle \eta^i|\,\,=\,\,
\frac1N\sum_i M^{-1} |\eta^i\rangle\langle {\eta^i}| \quad\,\,=\,\,
\quad M^{-1}\left[\mathds{1}+O\left(1/\sqrt{N}\right)\right]\,. 
\label{estimate_eq}
\end{equation}

We obtain the full propagator plus off-diagonal noise term,
which vanish like $\frac{1}{\sqrt{N}}$.
We apply several improvement schemes to reduce this stochastic noise.

\subsection{Staggered spin dilution (SSD)}
Usually a propagator receives its dominant noise contributions
from the stochastic sources that are closest to the sink,
in terms of Euclidean distance or in spinor space.
Such contributions can be reduced in a straight forward way by 
{\it partitioning} (or ``diluting'') spacetime, colour and/or spin
into $n_p$ disjoint subspaces~\cite{Bernardson:1993yg,Foley:2005ac}.
The subsequent reconstruction of an all-to-all propagator
from partition-to-all propagators comes at a
significant computational overhead (proportional to 
$n_p$). Sometimes the resulting statistical error
however is reduced by more than a factor $\propto n_p^{-1/2}$, justifying
this method.
 
\begin{figure}[!ht]
\begin{tabular}{ccccS}
 {\bf Scheme} & & & & {\bf Nearest neighbor coupling} \\
\\
 \begin{tabular}{c|c|c|c|c|c|c|c}
1 & 1 & 1 & 1 & 1 & 1 & 1 & 1 \\
\hline
1 & 1 & 1 & 1 & 1 & 1 & 1 & 1 \\
\hline
1 & 1 & 1 & 1 & 1 & 1 & 1 & 1 \\
\hline
1 & 1 & 1 & 1 & 1 & 1 & 1 & 1 \\
\hline
... & & & & & & & 
\end{tabular}
& & & &
\resizebox{70pt}{!}{\includegraphics[clip]{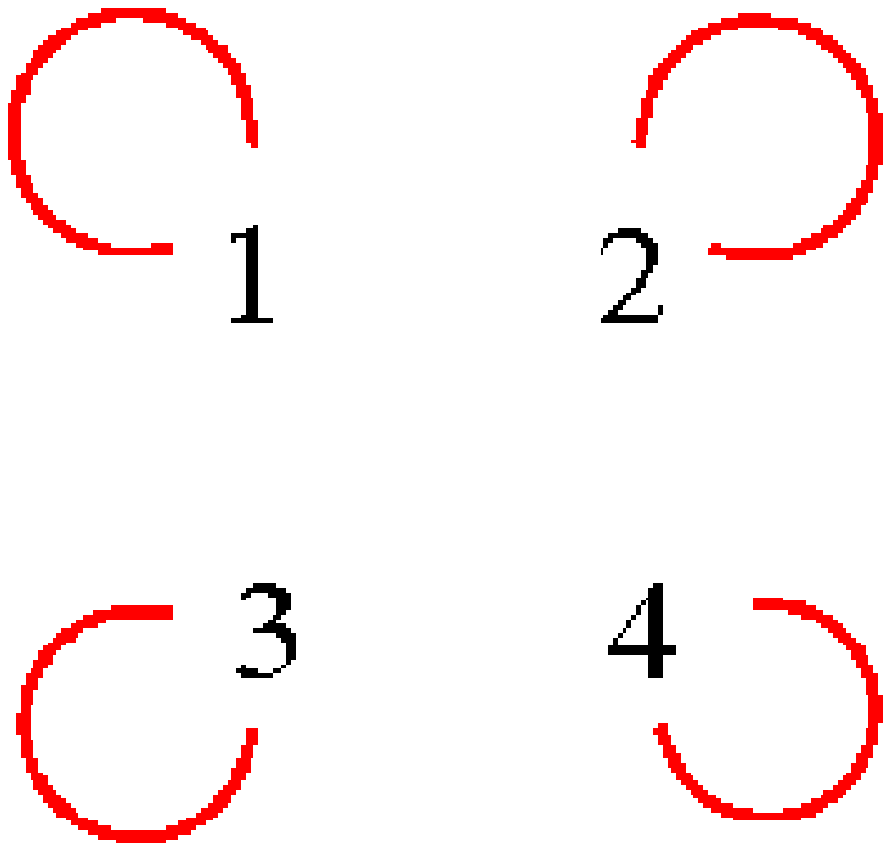}} \\
\\
\begin{tabular}{c|c|c|c|c|c|c|c}
1 & 3 & 2 & 4 & 1 & 3 & 2 & 4 \\
\hline
3 & 2 & 4 & 1 & 3 & 2 & 4 & 1 \\
\hline
2 & 4 & 1 & 3 & 2 & 4 & 1 & 3 \\
\hline
4 & 1 & 3 & 2 & 4 & 1 & 3 & 2 \\
\hline
... & & & & & & &  
\end{tabular}
& & & &
\resizebox{40pt}{!}{\includegraphics[clip]{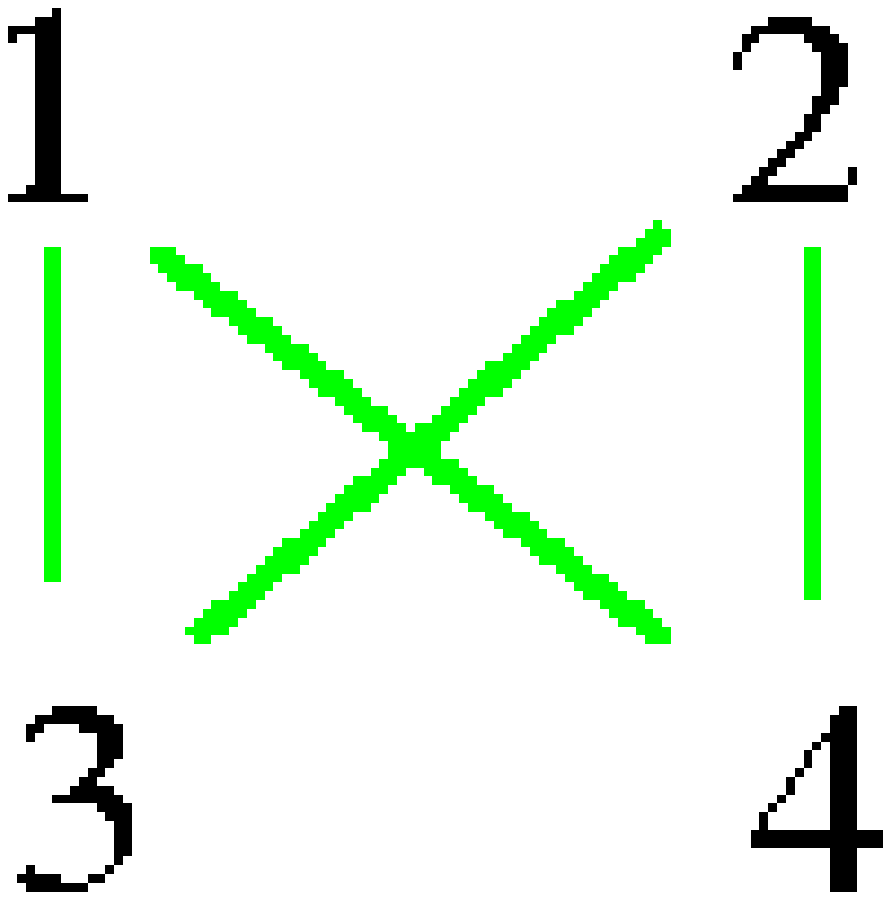}} 
\end{tabular}
\caption{The left column shows the dilution schemes in two dimensions. The corresponding coupling is sketched in the right column (green is weak, red is strong coupling).\label{SSD_fig}}
\end{figure}

Spin partitioning is usually achieved by employing four sets
of noise vectors, each with only one spin component different from zero.
Obviously, this is not the only possible pattern. 
We find that especially for heavy quarks other distributions
of the spin components are superior to this standard scheme.
We attempt to exploit the fact that for large quark masses
the Wilson-Dirac operator only weakly couples the upper and
the lower spinor components
and we devise a partitioning
pattern where at different spacetime positions
different spin components
are stochastically seeded: an upper component site should only
neighbour lower component sites and vice-versa.
We found a scheme to be particularly effective,
which we coin {\it Staggered spin dilution (SSD)}.
This is defined and compared to the standard dilution scheme in the
left column of Fig.~\ref{SSD_fig}.
The numbers indicate the spinor component which is different
from zero. The right column of Fig.~\ref{SSD_fig} shows
the corresponding nearest neighbour coupling strength,
where red means ''strong'' and green means ''weak''. 

\subsection{Hopping parameter acceleration (HPA)}
This technique~\cite{Thron:1997iy,Wilcox:1999tc,Bali:2005fu} is based on
the {\it Hopping Parameter Expansion (HPE)}
of a Wilson-type Dirac operator
$M=\mathds{1} - \kappa D$, where $D$ only couples nearest neighbours: 
\begin{eqnarray}
M^{-1} & = &
(\mathds{1}-\kappa D)^{-1} =
 \mathds{1}+\kappa D + \ldots + (\kappa D)^{n-1}+
(\kappa D)^n\sum_{i=0}^{\infty}(\kappa D)^i \\
& \Rightarrow & (\kappa D)^n M^{-1} = M^{-1} - ( 1+\kappa D + \ldots + \kappa D^{n-1} ) \,:
\end{eqnarray}
the multipication of the propagator by $(\kappa D)^n$
subtracts all the terms 
up to $n-1$ hops, cancelling the corresponding
fluctuations from noisy
source positions near the sink. While this reduces the
variance, the signals of Green functions for distances
larger than $n-1$ hops remain unaffected.
Applications are the calculation of
correlation functions with time separations $t\geq na$
or of disconnected loops where the first $n-1$ terms are either
calculated exactly or vanish identically. The effect of HPA
on the pseudoscalar loop calculated using the Wilson action
at the charm mass is plotted on the left hand side of Fig.~\ref{HPA-RNS_fig}.
(The first multiplication has no effect since
$\mathrm{Tr}\,\gamma_5$ is zero anyway.)
This technique is particularly suitable for heavy quarks
where the rate of convergence of HPE improves but
significant gains have also been reported for light quarks~\cite{Collins:2007mh}.

\subsection{Recursive noise subtraction (RNS)}
\begin{figure}
\parbox[l]{177pt}{
 \resizebox{177pt}{!}{\includegraphics[clip]{kD.eps}}
}
\parbox[c]{30pt}{\mbox{}}
\parbox[r]{219pt}{
\vspace{-0.25cm}
 \resizebox{170pt}{!}{\includegraphics[clip]{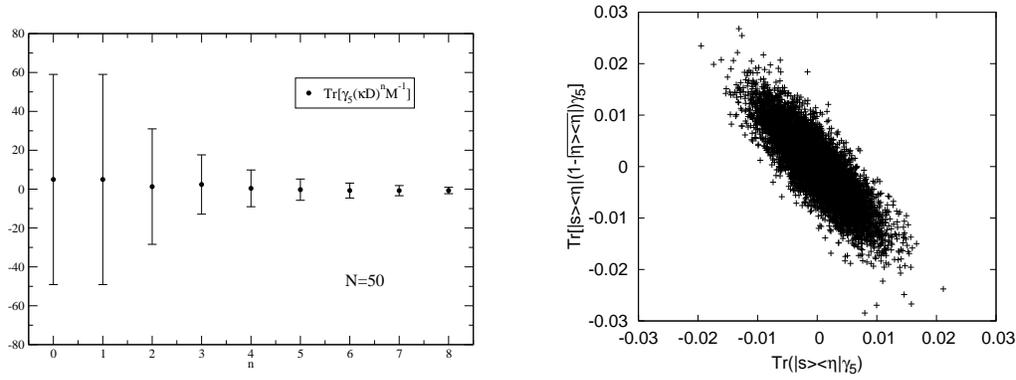}}
}
\caption{The effect of the HPA on the error of the pseudoscalar
correlator for the Wilson action at the charm mass is shown on the left.
On the right the RNS scatter plot is displayed.}
\label{HPA-RNS_fig}
\end{figure}

In addition we use an algebraic improvement scheme
(RNS) where the off-diagonal
terms appearing in Eq.~(\ref{estimate_eq}) are calculated and
subtracted by hand. The idea can be illustrated as follows:
\begin{equation}
 M^{-1} \,\, =\,\,  \overline{|s\rangle \langle \eta|} \ + \ M^{-1} (1-\overline{|\eta \rangle \langle \eta|})
 \,\,\approx\,\,  \overline{|s\rangle \langle \eta|} \ + \ \overline{|s\rangle \langle \eta|} (1-\overline{|\eta \rangle \langle \eta|})\,.
\end{equation}
The outer product $\overline{|\eta\rangle\langle\eta|}$
should be truncated to diagonal blocks of a dimension that is small
compared to $\sqrt{N}$. We use the 12-dimensional  colour $\otimes$ spin
subspace and display a scatter plot between estimates on the right
hand side of Fig.~\ref{HPA-RNS_fig}. The correlation angle is close to
$\pi/4$: adding the two terms as suggested will result in a reduced
stochastic error. In principle one can make a more
general ansatz like
$\overline{|s\rangle \langle \eta|} [\mathds{1}+\alpha
(\mathds{1} -\overline{|s\rangle \langle \eta|})
+\beta(\mathds{1} -\overline{|s\rangle \langle \eta|})^2+\cdots]$
and optimize the parameters $\alpha, \beta,\ldots$ accompanying the
different estimates of {\em zero} to minimize the noise in a given channel.
Here we use $\alpha=1, \beta=0$.

\subsection{Truncated solver method (TSM)}
For the computation of the light quark loops appearing in the
mixing matrix we apply the TSM
using the BiCGStab solver.
For details see \cite{Collins:2007mh}.

\subsection{Reduction of the total computational effort}
We summarize the effect of the tested improvement
schemes on the
disconnected part of the zero-momentum projected
$\eta_c$ two-point function
$\langle \ \mathrm{Tr}(M^{-1}\gamma_5) \ \mathrm{Tr}(M^{-1}\gamma_5) \ \rangle$
at $t=0$ in  Table \ref{nr_tab}. The gain is in terms of real computer time.
$n$ is the power of $\kappa D$ applied to the sink vector.
Our use of the clover action to calculate $\mathrm{Tr}(M^{-1}\gamma_5)$
restricts us to $n<3$, since $\mathrm{Tr}(D^2\gamma_5) \propto F \tilde{F}\neq 0$. In principle one could calculate these terms by hand
and add them back in again.
With colour and staggered spin dilution
and two applications of $\kappa D$ we obtain a net gain factor
of almost 12. Note that these numbers do not include the effect of
the TSM that we only use for the light quark propagators.
Also we have not yet combined RNS with the partitioning methods. 

\begin{table}[!ht]
\begin{center}
\begin{tabular}{c||c|c|c|c|c|c|c}
n & no & spin & color & color + spin & SSD  & SSD + color & RNS \\
\hline
\hline
0 & 1 & 1.43 & 1.80 & 2.52 & 1.97 & 3.63 & 1.87 \\
\hline
2 & 2.89 & 6.32 & 5.06 & 10.24 & 7.16 & 11.80 & 5.44  
\end{tabular}
\end{center}
\caption{Effective gain of the different improvement schemes. $n$ is the power of $\kappa D$ used in HPA.}
\label{nr_tab}
\end{table}

\section{The mixing}
\begin{figure}[!ht]
\begin{center}
 \resizebox{250pt}{!}{\includegraphics[clip]{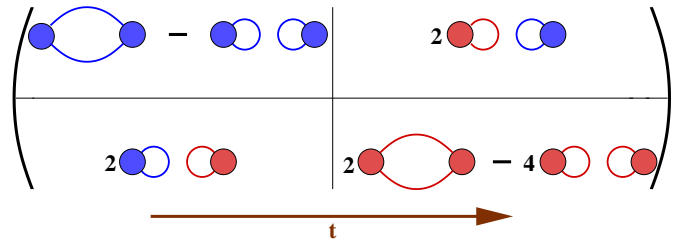}}
\end{center}
\caption{The mixing matrix. The blue lines indicate charm quark,
the red ones light quark propagators.}
\label{mixingmatrix_fig}
\end{figure}
Now that we are equipped with techniques to address
quark-antiquark-annihilation diagrams reliably,
we calculate the elements of the mixing matrix. Both charmonium
and light meson interpolators are included, three in each sector:
$(c\bar{c})_{0}$, $(c\bar{c})_{10}$, $(c\bar{c})_{80}$,
$(u\bar{u})_{0}$,  $(u\bar{u})_{5}$, $(u\bar{u})_{40}$,
where the subscript indicates the number of Wuppertal smearing steps
with $\delta=0.3$, employing spatial APE smeared ($n_{APE}=15$, $\alpha=0.3$)
parallel transporters. 
These smearing parameters were determined by
optimizing the effective masses within the two sectors, ignoring
disconnected contributions, see \cite{Ehmann:2007hj}.
In Fig.~\ref{mixingmatrix_fig} we sketch the two by two mixing matrix,
where the blue lines represent charm and the red lines
light quark propagation.
The prefactors are due to the two mass degenerate light flavours.

\begin{figure}[!ht]
\parbox[l]{200pt}{
 \resizebox{200pt}{!}{\includegraphics[clip,angle=270]{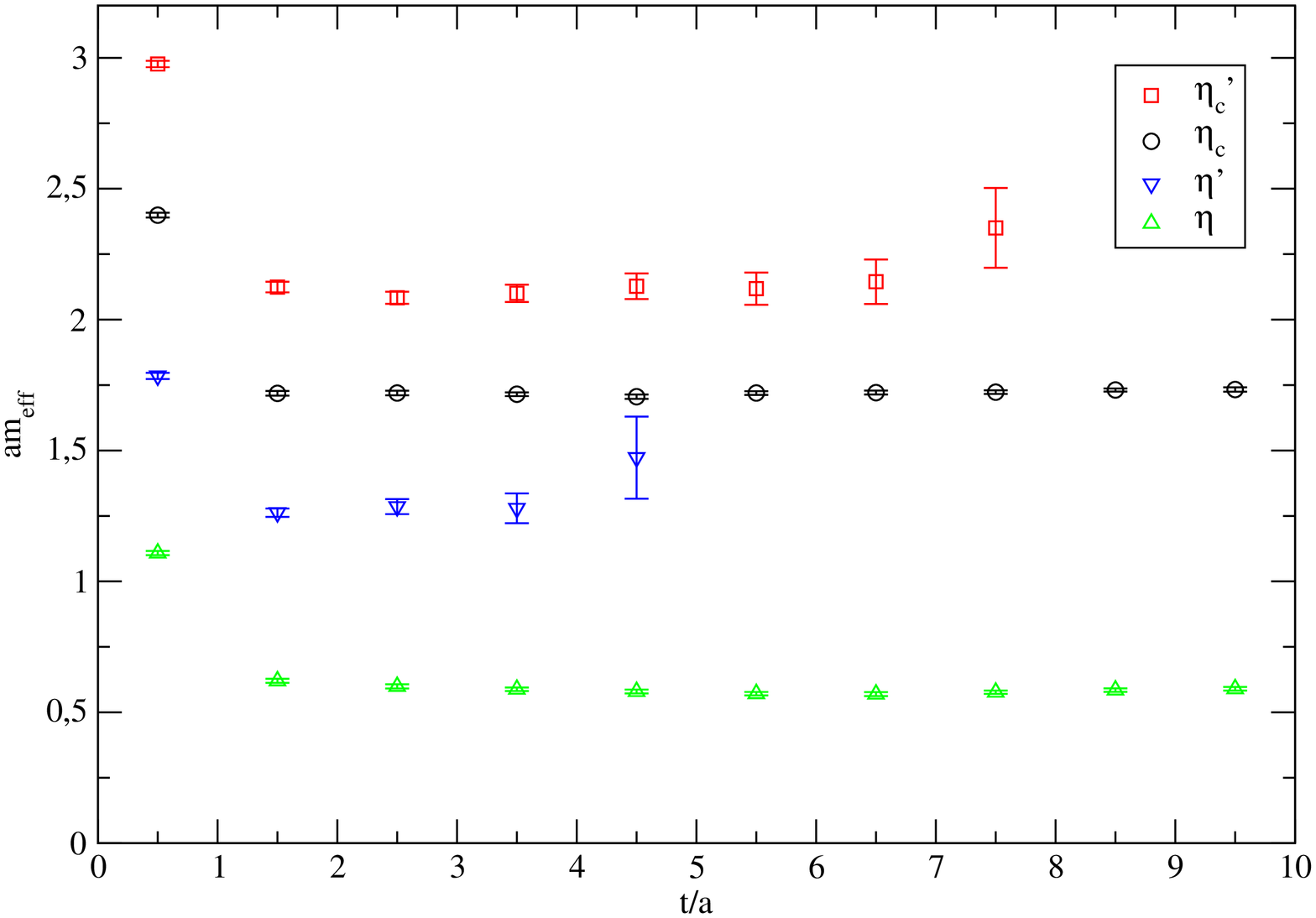}}
}
\parbox[c]{20pt}{\mbox{}}
\parbox[r]{200pt}{
 \resizebox{200pt}{!}{\includegraphics[clip,angle=270]{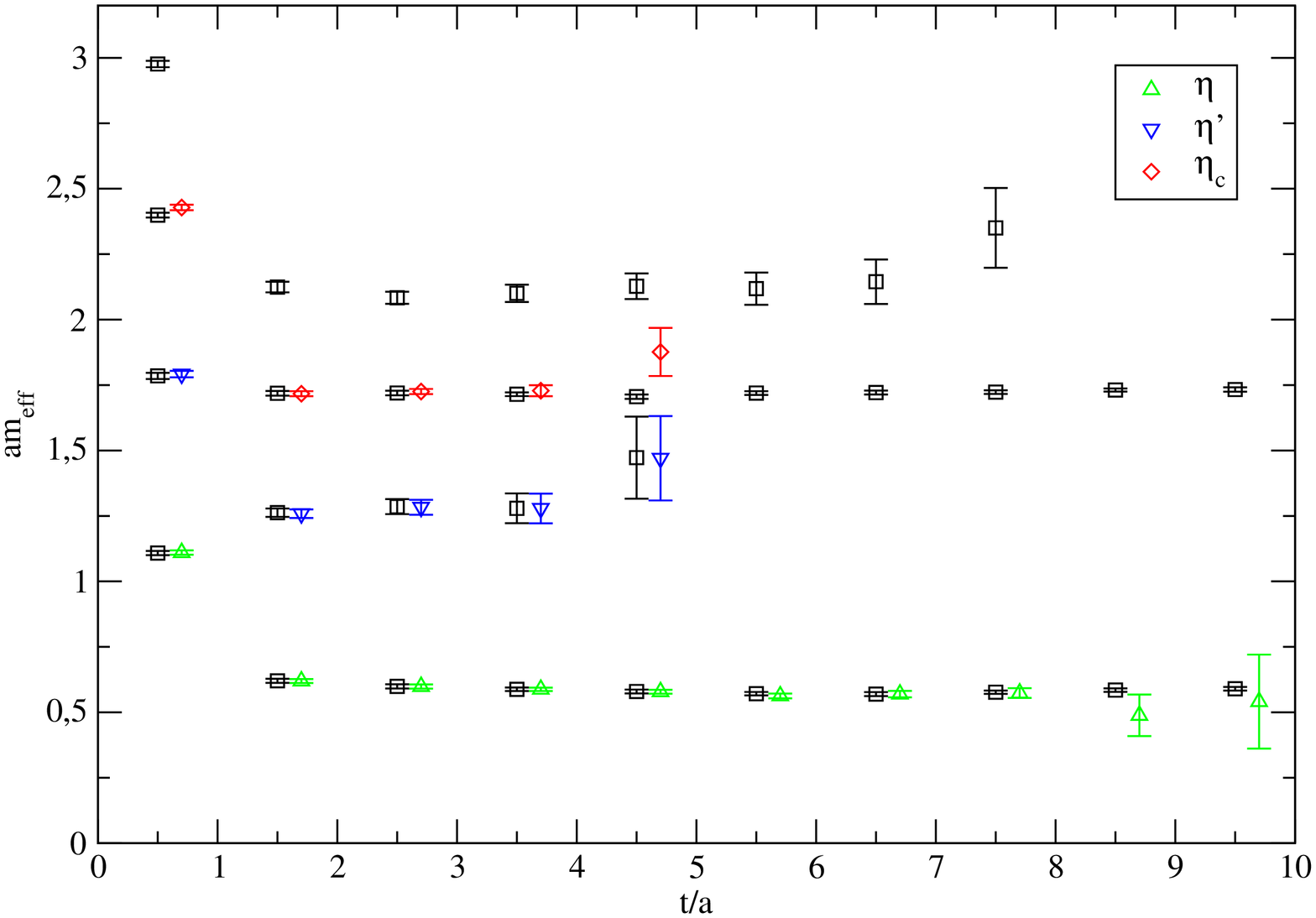}}
}
\caption{The left plot shows the effective masses of the eigenvalues obtained from diagonalizing the corresponding submatrices. To the right
these eigenvalues (in black) are plotted together with the ones obtained from the basis $(c\bar{c})_{10}$, $(c\bar{c})_{80}$, $(u\bar{u})_{5}$, $(u\bar{u})_{40}$.} 
\label{em_fig}
\end{figure}

\begin{figure}[!ht]
\parbox[l]{200pt}{
 \resizebox{200pt}{!}{\includegraphics[clip,angle=270]{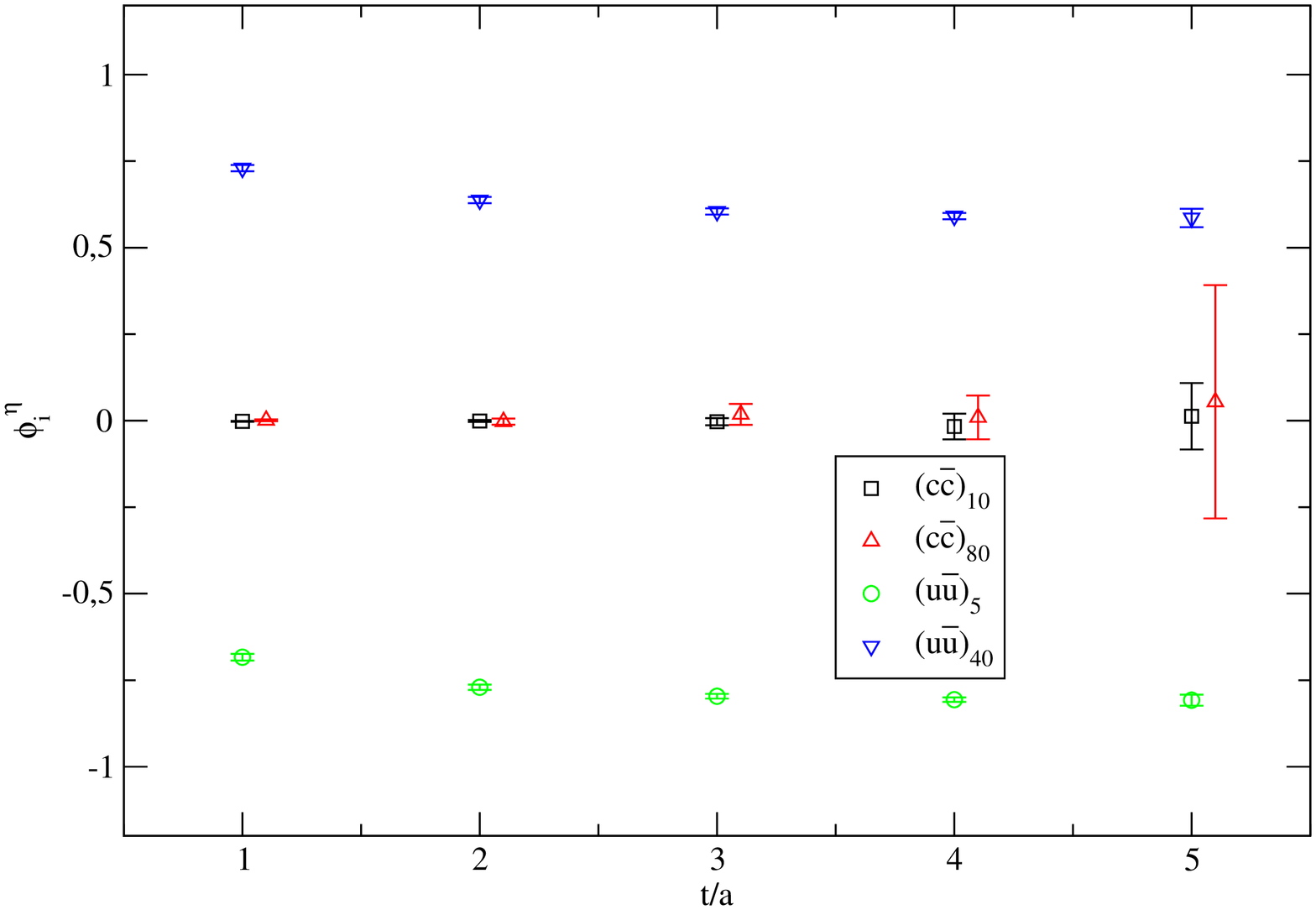}}
}
\parbox[c]{20pt}{\mbox{}}
\parbox[r]{200pt}{
 \resizebox{200pt}{!}{\includegraphics[clip,angle=270]{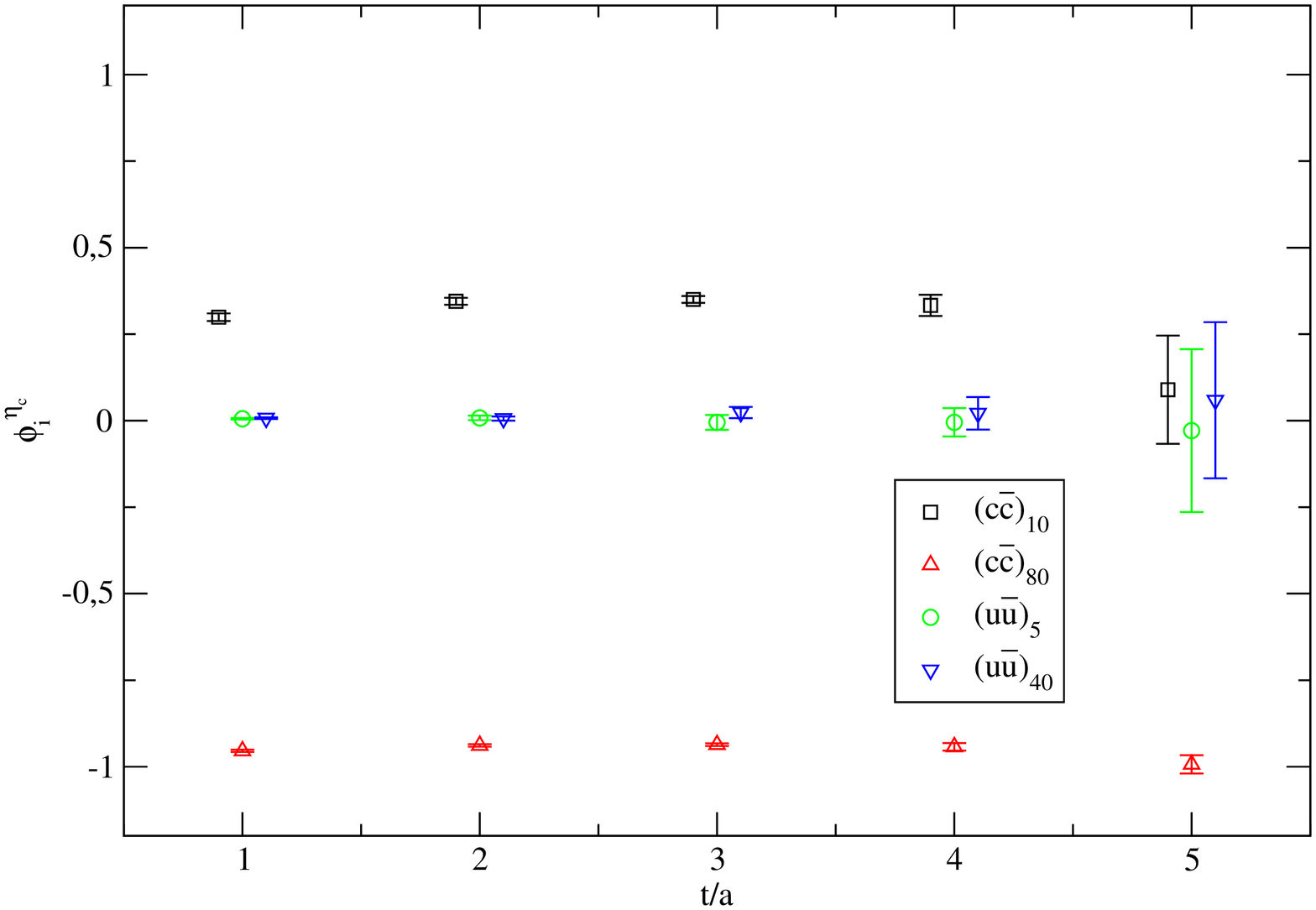}}
}
\caption{The eigenvector components of the $\eta$ are displayed on
the left, the ones of the $\eta_c$ on the right.}
\label{evec_fig}
\end{figure}

The variational method is applied to the mixing matrix by
solving a generalized eigenvalue problem,
$C(t_0)^{-1/2} \, C(t) \, C(t_0)^{-1/2} \, \psi^{\alpha} =
\lambda^{\alpha}(t,t_0) \, \psi^{\alpha}$, see \cite{Ehmann:2007hj}
for details. For sufficiently large times the eigenvalues
and -vectors will approach their asymptotic values.
The components of a given eigenvector
can be interpreted as the coupling strengths of the
corresponding operators to
the state under consideration.
Their magnitude will provide us with information about the magnitude
of mixing in the system. 
We proceed as follows: we first determine the
eigenvalues of the 3 by 3 submatrices separately
within each of the flavour sectors,
where for the moment being we ignore the disconnected contribution
in the charmonium sector (see the left hand side of Fig.~\ref{em_fig}).
The light $\eta$ and its first radial excitation ($\eta'$) are
the lowest two eigenvalues of the submatrix containing only 
light interpolators and $\eta_c$ and $\eta_c'$ within the charmonium
sector. We find a diagonalisation of the full 6 by 6 matrix
to be numerically unstable
and hence restrict ourselves to the basis
of the states $(c\bar{c})_{10}$, $(c\bar{c})_{80}$,
$(u\bar{u})_{5}$ and $(u\bar{u})_{40}$ for the full-fledged
mixing analysis.
In the right hand side of Fig.~\ref{em_fig}
the effective masses of the lowest three eigenvalues
obtained from this basis are shown together with the eigenvalues obtained
above, ignoring the mixing effects: within statistical errors
no effect is seen.
The eigenvectors contain more detailed information about the mixing.
We display the components of the ground state $\eta$ eigenvector
on the left hand side of Fig.~\ref{evec_fig} and those of the
$\eta_c$ eigenvector on the right hand side.
Indeed, the $\eta$ does not contain any statistically significicant
admixture from the $\eta_c$ sector and vice versa.

We conclude that there is no significant $\eta$-$\eta_c$-mixing and neither
are there any other significant flavour singlet effects on the $\eta_c$ mass.
However, the unrealistically heavy pion mass might have affected
our conclusion. Runs on $24^3\times 48$ lattices with
$m_{\pi} \approx 400$ MeV will clarify this issue. Furthermore,
mixing with other states like glueballs deserves future attention.

\vspace{1cm}

\acknowledgments
This work was supported by the
BMBF (contract 06RY257, GSI-Theory).
We thank the DFG 
Sonderforschungsbereich/Transregio 55
for their support.

\end{document}